# Ludvig Lorenz, Electromagnetism, and the Theory of Telephone Currents

Helge Kragh[*]

**Abstract**: Ludvig V. Lorenz (1829-1891) was Denmark's first theoretical physicist and the only one whose work attracted international attention in the second half of the nineteenth century. This paper presents a survey of Lorenz's contributions to physics with an emphasis on his work in electrodynamics and electrical science. His 1867 electrodynamic theory of light was of a theoretical and foundational nature, while his unpublished theory of telephone currents was oriented toward practical problems in long-distance telephony. Lorenz's theories are briefly compared to those of better known physicists such as H. A. Lorentz, J. C. Maxwell, and O. Heaviside.

The Danish nineteen-century physicist Ludvig V. Lorenz is not well known today (and is sometimes confused with the famous Dutch physicist Hendrik A. Lorentz). Yet his life and contributions to physics are described in, for example, the *Dictionary of Scientific Biography* and several other sources in the history of science literature [e.g. Pihl 1939; Pihl 1972; Kragh 1991]. A French translation of Lorenz's collected papers was published by the mathematician Herman Valentiner in two valuable but somewhat obscure volumes [Valentiner 1898-1904]. Leading physicists in the second half of the nineteenth century were aware of his work and often referred to it. James Clerk Maxwell, Ludwig Boltzmann, Heinrich Hertz, Friedrich Kohlrausch, and Carl Anton Bjerknes were among those who either corresponded with Lorenz or referred to his publications.

In this paper I give a summary account of Lorenz's work in pure and applied electrodynamics, in particular his electromagnetic theory of light and his attempt to design on a theoretical basis telephone cables for communication over long distances. Some of his other work will also be mentioned, but only rather briefly. Lorenz was in several respects a representative of the new theoretical physics which emerged in the late nineteenth century [Jungnickel and McCormmach 1986]. Like other physicists of

[*] Niels Bohr Institute, University of Copenhagen, Denmark. E-mail: helge.kragh@nbi.ku.dk. The paper is based on a lecture given at the International Conference on the History of Physics held at Trinity College Cambridge in September 2014.

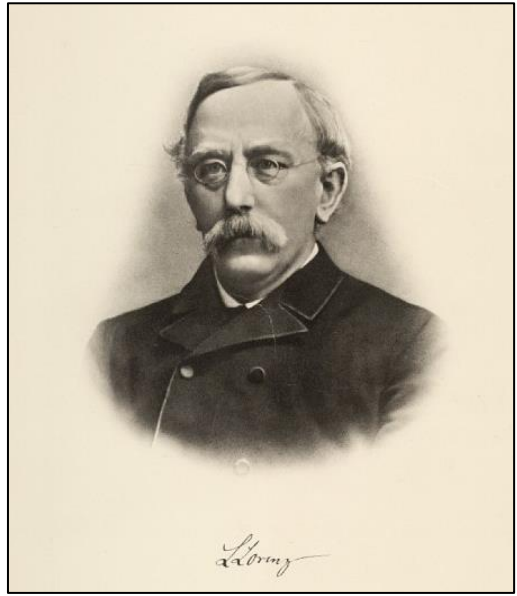

Ludvig V. Lorenz (1829-1891)

his generation he applied his mathematical skills not only to theoretical problems of physics but also to practical problems of a technical nature.

## 1. Biographical sketch

Born in Elsinore, Denmark, in 1829, Ludvig Valentin Lorenz was early on fascinated by mathematics and physics. As he recalled in an autobiographical note: "An evening lecture in physics during the winter [of 1842] aroused my interest in the field at an early date, and I soon realized that I wanted to make the study of mathematics and physics my calling; this goal was constantly in my mind." The lecture that made such an impression on 13-year-old Lorenz was arranged by the Society for the Dissemination of Natural Science, an organization that H. C. Ørsted had founded in 1824 and in part modelled after the Royal Institution in England [Kragh et al. 2008, p. 178].

However, Lorenz's road toward a career as a physicist was anything but smooth. As a young man he attended a variety of courses at the Polytechnic College and the University of Copenhagen, finally graduating as a polytechnic candidate in chemistry in 1852. He never received a formal education in physics, a field he much preferred over chemistry. Supported by government and private grants, in 1858 Lorenz was able to go to Paris to improve his knowledge of theoretical physics. While in Paris he specialized in theories of optics and elasticity, which subjects he continued to cultivate after his return to Copenhagen. In 1866 he was elected a member of the prestigious Royal Danish Academy for Sciences and Letters and the same year he was appointed teacher at the Military High School just outside Copenhagen. He stayed in this humble position for 21 years, unable to obtain a position at either the University or the Polytechnic College. Eventually, in 1887 he accepted a most generous offer from the Carlsberg Foundation which J. C. Jacobsen, the wealthy owner of the Carlsberg Breweries, had established nine years earlier. The Carlsberg Foundation offered to pay Lorenz as an independent researcher for the rest of his life. He died only four years later by a heart attack.

While much of Lorenz's scientific work was within the fields of electrodynamics and optics, he also made contributions to other branches of physics. Thus, during the 1870s he studied the relationship between the electrical and thermal conductivities of metals. In 1872 he concluded from experiments that a metal's specific heat conductivity λ was related to its heat conductivity σ by the formula

$$\frac{\lambda}{\sigma} \cong \text{const} \times T \,,$$

where $T$ is the absolute temperature. This law was a generalization of the empirically based Wiedemann-Franz law of 1853 which did not take into account the temperature variation. The Wiedemann-Franz-Lorenz law would later play an important role in the electron theory of metals. In his doctoral dissertation of 1911, a critical and innovative review of the electron theory of metals, young Niels Bohr referred to Lorenz's paper and how the law had later been explained by Paul Drude, H. A. Lorentz and other electron theorists.

## 2. Contributions to optical theory

Having initially adopted the elastic or mechanical models of light favoured by Cauchy and other French physicists, from the early 1860s onwards Lorenz advocated a purely phenomenological theory that was independent of the structure of matter. He wanted to base the theory of light solely on directly observable quantities such as intensity, velocity, direction of propagation, colour, phase, and polarization plane. On the other hand, he did *not* count the luminiferous ether on which most conceptions of light were based as belonging to this class. According to Lorenz, the ether was a dispensable construct with no real existence in nature. In a Danish paper of 1867 he dismissed the ether as a "non-substantial medium which has been thought of only because light was conceived in the same manner as sound and hence there had to be a medium of exceedingly large elasticity and small density in order to explain the large velocity of light." To his mind the ether was blatantly *ad hoc*.

Lorenz's best known work in optics resulted from a systematic study of refraction and its relation to the molecular state of the transparent medium [Lorenz 1880]. On the basis of his phenomenological theory of light he derived in 1869 a relationship between the refractive index $n$ (for wavelength $\lambda \to \infty$) and the density $d$ of the medium:

$$\frac{n^2 - 1}{n^2 + 2} = const \times d$$

Lorenz showed that the quantity on the left side, the refractivity, is additive and thus valid also for mixtures of gases or liquids. If the mixture consists of $r$ components, the law reads

$$\sum_{i=1}^{r} \frac{m_i^2 - 1}{(m_i^2 + 2)d_i} = const$$

Realizing that the refractivity depends on the medium's molecular structure Lorenz developed his theory into a method for estimating the size of molecules. In 1890 he derived in this way values for the number of molecules in a unit volume of air (Avogadro's number $N_A$) and the radius $R$ of the molecules:

$$N_A \cong 3.7 \times 10^{23} \text{ molecules/litre}$$

and

$$R \geq 1.4 \times 10^{-10} \text{ m}$$

Unaware of the Copenhagen physicist, 25-year-old H. A. Lorentz deduced the same refraction formula in 1878, although on the basis of Helmholtz's electromagnetic theory of light. He later referred to the "Lorenz-Lorentz" formula as "a curious case of coincidence" [Lorentz 1909, p. 145]. Since Lorentz published his paper in Dutch, and Lorenz published his in Danish, the refraction formula became generally known only when German translations of their papers appeared in *Annalen der Physik* in 1880. Then the formula spurred a large number of experiments in molecular refractivity and generally became an important tool in the new physical chemistry. Studies of molecular refractivity were eagerly taken up by H. H. Landolt, W. Ostwald, J. W. Brühl and other chemists. By applying this method they obtained valuable information concerning the bonds in chemical compounds, their constitution, and the size of molecules and ions [Nernst 1904, pp. 306-13]. Lorenz did not partake in this work and his later attempt to develop his refraction theory into a theory of dispersion was unsuccessful. The main reason was that his dispersion formula was unable to account for the anomalous dispersion discovered by his compatriot Christian Christiansen in 1870.

The other area of optics in which Lorenz is still remembered today is the theory of scattering of light by small spherical bodies. This mathematically complex problem was attacked by Lorenz in the mid-1880s and resulted in 1890 in an extensive paper in the proceedings of the Royal Danish Academy. Written in Danish it remained unknown to most physicists even after a French translation appeared in Lorenz's collected papers. Much of the theory of Rayleigh scattering, published by Lord Rayleigh in 1899, can be found in Lorenz's earlier paper [Kerker 1969, pp. 27-39; Kragh 1991]. Likewise, the important theory of plane wave scattering published by Gustav Mie in 1908 was to some extent anticipated by Lorenz's work [Logan 1965; Mishchenko and Travis 2008]. According to some modern physicists, the Mie theory should properly be called the Lorenz-Mie theory.

Lorenz also applied his optical scattering theory to explain the rainbow from the fundamental principles of optical physics. At the end of his life he prepared a detailed manuscript of the theory of the rainbow, but the extensive calculations were never published. To this day they remain in his archival deposits.

## 3. Retarded potentials and the Lorenz gauge

In 1867 Lorenz published an important paper on electrodynamics that appeared in both German (*Annalen der Physik und Chemie*) and English (*Philosophical Magazine*). Probably inspired by his former teacher H. C. Ørsted, the discoverer of electromagnetism and professor at the Polytechnic College, he aimed at unifying the forces of nature without introducing new physical hypotheses. "The idea that the various forces in nature are merely different manifestations of the one and same force has proved itself more fertile than all physical theories," he wrote [Lorenz 1867, p. 287]. As to the physical theories or hypotheses, they "have only been useful inasmuch as they furnish a basis for our imagination." Lorenz was referring to the ether, which "there is scarcely any reason for adhering to." As an alternative to the ether he ascribed a non-zero electrical conductivity to empty space: "It may well be assumed that in the so-called vacuum there is sufficient matter to form an adequate substratum for the motion."

The introduction to Lorenz's 1867 paper in *PhilosophicalMagazine*

The novel element in Lorenz's theory, which built on potentials rather than fields, was his introduction of retarded potentials instead of the instantaneous electrical action assumed by German physicists such as Robert Kirchhoff and Franz Neumann. Retardation is necessary, he emphasized, in order to account for the propagation of electromagnetic disturbances in general and for light in particular. By the 1860s it was widely accepted that the electromagnetic field could be expressed in terms of the scalar potential φ and the vector potential **A**. The idea of using φ in a retarded form had been proposed by Bernhard Riemann in 1858, but only in an oral presentation and without including the vector potential. Lorenz was the first to propose the full retarded potentials, which in modern standard notation can be written as

$$\varphi(\boldsymbol{r}, t) = \int \frac{\rho(\boldsymbol{r}', t - R/c)}{R} dV'$$

$$\boldsymbol{A}(\boldsymbol{r}, t) = \int \frac{\boldsymbol{j}(\boldsymbol{r}', t - R/c)}{R} dV'$$

$R$ denotes the distance between the points **r** and **r'**, and $\rho$ and $j$ are the charge density and current density, respectively. On this basis and by adding a term corresponding to Ohm's law, Lorenz established a system of equations which was mathematically equivalent to Maxwell's field equations but differed from them in a physical sense.

In the course of deriving his equations Lorenz had to consider the condition fixing the vector potential. In a purely mathematical way he derived what is currently known as the Lorenz gauge condition, which in SI units can be written as

$$\nabla \cdot \boldsymbol{A} + \frac{1}{c^2} \frac{\partial \varphi}{\partial t} = 0$$

The condition is Lorentz invariant and thus satisfies Einstein's theory of relativity proposed nearly forty years later. Using the relativistic four-potential $A_\mu$ ($\mu = 0, 1, 2, 3$) it can be expressed in the more compact form

$$\partial_\mu A^\mu = 0$$

For a very long time H. A. Lorentz got credit for the "Lorentz condition," which can be considered an example of what sociologists of science call Stigler's law. According to this "law," a scientific discovery is never or only rarely named after its original discoverer [Stigler 1980; Jackson 2007]. Lorentz did extremely important work in electrodynamics, of course, but he was not in fact the originator of either the gauge condition or the general expressions of the retarded potentials [Jackson and Okun 2001; Nevels and Shin 2001].

　　Strangely, Lorenz never mentioned Maxwell's theory of electromagnetism. In 1867 he was unaware of it, but even later on, when he assumedly did know about it, he ignored the theory. On the other hand, Maxwell was aware of the theory of the Danish physicist to which he referred in a paper of 1868 and again in his famous *Treatise on Electricity and Magnetism* from 1873. From Maxwell's point of view, which assumed the so-called Coulomb gauge, namely

$$\nabla \cdot \mathbf{A} = 0,$$

there was no reason to take Lorenz's alternative seriously. He wrongly suggested that Lorenz's retarded potentials violated Newton's third law of motion as well as the law of energy conservation [Maxwell 1965, vol. 2, p. 137].

## 4. An electrodynamic theory of light

The specific aim of Lorenz's 1867 paper was to construct a theory of light on the basis of a form of electrodynamics that differed from both the British field view and the German action-at-a-distance view [Rosenfeld 1979; Darrigol 2000, pp. 211-213]. This he did by modifying Kirchhoff's equations so that the motion of electricity would become analogous to that of the optical ether. "Electrical forces require time to travel," Lorenz pointed out, "and every action of electricity and of electrical currents does in fact only depend on the electrical condition of the immediately surrounding elements." Moreover, the new theory had to comply in its mathematical aspects with his previous phenomenological theory of transverse light waves. Indeed, he showed that the wave equation of the previous theory reappeared almost identically as a wave equation of the electrical current density $\boldsymbol{j}$. This equation he wrote in a form that included the vacuum permeability $\mu_0$, the conductivity of free electricity σ, and an indeterminate constant $a$. Contrary to the Maxwell equations, the vacuum permittivity $\varepsilon_0$ had no place in Lorenz's theory.

We assume, Lorenz wrote in his 1867 paper, "that the vibrations of light themselves are electrical currents … while $a$ is the velocity with which electrical action is propagated through space." In vector notation his wave equation reads

$$-\nabla \times \nabla \boldsymbol{j} = \frac{1}{a^2} \frac{\partial^2 \boldsymbol{j}}{\partial t^2} + \mu_0 \sigma \frac{\partial \boldsymbol{j}}{\partial t}$$

Apart from the last term the equation was the same as the one he had used in his optical theory. The last term is an absorption term and according to Lorenz it explained the well-known relationship between conductivity and absorptive power implying that electrical conductors are opaque. Instead of speaking of the ether and of optical vibrations, he now spoke of a poor conductor and alternating electrical currents. The quantity $a$ obviously was of importance. Although its value was not given by the theory, Lorentz reasoned that it had to be of the order $c$, the velocity of light in vacuum, and that its square was likely to be equal to $c^2/2$. In that case,

$$a = \frac{c}{\sqrt{2}},$$

His equation would then "represent a mean between Weber's and Neumann's theories." Maxwell, on the other hand, assumed from his field theory that the speed of light was given by the relation

$$c = \frac{1}{\sqrt{\epsilon_0 \mu_0}}$$

The main result of Lorenz's investigation was "not only that the laws of light can be deduced from those of electrical currents, but that the converse way may be pursued." This result, he emphasized, "has been obtained without the assumption of a physical hypothesis, and will therefore be independent of one." Lorenz optimistically concluded that his equations, "lead us a step further towards developing the idea of the unity of forces, and open a fresh field for future inquiries." Giving his enthusiasm one might think that Lorenz followed up on his system of equations, but in fact he never returned to his electrodynamic theory of light or any of the "future inquiries" it allegedly opened up for. He may have thought that his theory was complete, or perhaps he did not fully realize its implications.

For a brief period of time Lorenz's theory was fairly well known among physicists in continental Europe and abroad. For example, it is known that Josiah Willard Gibbs studied it during his trip to Europe in the late 1860s. But it soon fell into oblivion, overshadowed by the more successful versions of electromagnetism developed by Maxwell and Helmholtz in particular. Lorenz's inexplicable decision not to develop or promote his theory undoubtedly also played a role. The most notable difference between the theories of Maxwell and Lorenz, apart from the ether, was that according to the latter theory light propagation was an expression of conduction currents. There was no analogy to Maxwell's displacement currents occurring in perfect insulators and therefore no possibility of extending Lorenz's theory to non-conducting media. In a certain sense Lorenz's theory was not electro*magnetic*, since it did not operate with the magnetic field.

The work that made Lorenz best known among the "electricians" of the Victorian era was not his electromagnetic theory of light but rather a clever experimental method that he first devised in 1873 to measure the ohm unit of resistance in an absolute way. The "absolute definition" of the ohm unit was a problem of international concern at the time. In England it was discussed by Lord Rayleigh who strongly advocated Lorenz's induction method as superior to other methods because it was free of the influence of terrestrial magnetism [Sidgwick and Rayleigh 1883]. Rayleigh's modification of the Lorenz method led to the adoption of a definition that expressed the so-called British Association unit as $0.98677 \times 10^9$ times the CGS unit. The method was later developed into the industrial standard for resistance used in Great Britain [Smith 1914].

## 5. The problem of long-distance telephony

In the decade from about 1878 to 1887 Lorenz turned towards technical aspects of physics, in particular electrical technologies related to the dynamo and the telephone. Collaborating with the Copenhagen instrument maker C. Jürgensen he designed new types of dynamos and electro-motors which were exhibited at the Paris electrical world exhibition in 1881 and also at a similar exhibition in Vienna in 1883. At about the same time he used the new telephone as a measuring apparatus in experiments with alternating currents. A few years later he began to study how a telephone current propagates in a metallic wire.

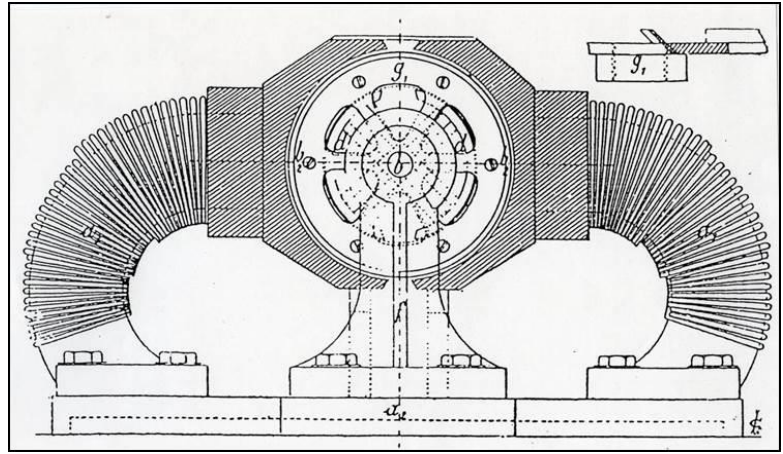

The Lorenz-Jürgensen dynamo.

This was more than just a matter of scientific curiosity, for it was also a problem of great economic significance for the young telephone industry in its attempt to compete with the efficient and well-established system of wire and cable telegraphy. In a nutshell, the problem was that for long or even medium distances of a few hundred km, the speech signals became so weak that they were barely audible. In addition the received signals were distorted, making it even harder to understand the transmitted message [Kragh 2009].

In trying to extend the range of telephone communication engineers tended to follow the tested rules of telegraphy in their design of cables and wire systems. The basic rule, which had its roots in the telegraph theory of William Thomson (Lord Kelvin), was that in order to reduce the attenuation the product of the line's capacitance $C$ and resistance $R$ per unit length should be minimized. With regard to self-induction ($L$) most engineers tended to ignore the factor or to consider it a kind of alternating-current resistance that increased the attenuation. Consequently $L$ should be as low as possible. According to William Preece, the later chief engineer of the British Post Office, "self-induction is a serious hindrance to rapid telegraphy and to long-distance speaking on telephones" [Preece 1887, p. 373]. The standard means of reducing the assumed harmful effect of self-induction was to use copper instead of iron in the wire and cable system.

Using the empirical rules of the engineers the speaking distance was actually improved, if not drastically. For example, the 498-km Paris-London line that opened for traffic in 1891 was considered a major technological achievement and a triumph of engineering methods in telephony [Tucker 1974]. The line used copper conductors of heavy gauge and the submarine Channel cable was supplied with heavy layers of guttapercha. However, it was soon realized that economic factors, meaning the price

Excerpt of Lorenz's manuscript on telephone theory.

of copper and guttapercha, put a limit on this kind of extension of the range of telephony. Practical and economically feasible telephone lines of length 1,000 km or more were not possible if based on the traditional methods inherited from the telegraph system. A few physicists and visionary engineers of a mathematical inclination argued that the very foundation of the design of telephone lines was wrong and that the main error was to think of telephone currents as a kind of rapidly alternating telegraph currents. In their view, the propagation of telephone currents was a field requiring its own scientific basis.

## 6. The Lorenz-Heaviside solution

This scientific basis was developed by the brilliant and eccentric British physicist Oliver Heaviside in a series of papers published between 1887 and 1890 [Heaviside 1892; Nahin 1988; Yavetz 1995]. A devoted Maxwellian, Heaviside based his theoretical analysis of the propagation of telephone currents on Maxwell's theory of electromagnetism, which still around 1890 was considered an arcane and academic theory in the world of practical engineers. In France, Aimé Vaschy independently arrived at basically the same results as Heaviside, but slightly later and without relying to any extent on Maxwell's theory [Atten 1988].

Even before Heaviside and Vaschy, Lorenz analysed theoretically how the transmission of the rapidly varying telephone currents took place in metallic wires. However, he never published his theory which is only known from a left manuscript probably dating from late 1885 [Kragh 1992]. Lorenz considered a single harmonic electrical wave of frequency ω propagating with velocity

$$v = \omega/\alpha$$

and with amplitude

$$I(x) = I_0 \exp(-\beta x)$$

While α is a velocity constant, β is an attenuation constant. Lorenz then set himself the task of expressing α and β in terms of the primary line constants, namely resistance (*R*), self-induction (*L*), and capacitance (*C*), all per unit length. Assuming the conductor to be perfectly insulated, he ignored the leakage constant *G*. By a straightforward series of mathematical calculations and making use only of Kirchhoff's and Ohm's laws, he ended up with the following results for the two secondary constants:

$$\alpha = \left(\frac{\omega C}{2}\left[\sqrt{R^2 + \omega^2 L^2} + \omega L\right]\right)^{\frac{1}{2}}$$

$$\beta = \left(\frac{\omega C}{2}\left[\sqrt{R^2 + \omega^2 L^2} - \omega L\right]\right)^{\frac{1}{2}}$$

As Lorenz pointed out, since a speech signal consists of many harmonic waves it will be distorted because the different components propagate with different velocities. Moreover, the attenuation of the amplitudes of the waves will differ according to their frequencies. Whereas Lorenz did not consider the leakage factor *G* in his manuscript of around 1885, he did so in other notes which are no longer extant [Kragh 1992, p. 316]. With this extension he expressed the results as

$$\alpha, \beta = \left[\frac{1}{2}\sqrt{(R^2 + \omega^2 L^2)(G^2 + \omega^2 C^2)} \pm (RG - \omega^2 LC)\right]^{\frac{1}{2}}$$

The + sign refers to the attenuation factor α, and the − sign to the distortion factor β.

The attenuation problem was the most important. Lorenz fully realized that in general increased self-inductance would make β smaller, that is, increase the speaking distance. He showed that in the case of ω*L* >> *R* and *L* not too great, the expression for β could be approximated to

$$\beta \cong \frac{R}{2}\sqrt{\frac{C}{L}}$$

In other words, theoretically the speaking distance would increase if self-inductance was *added* to the line and *R* and *C* kept as small as possible. To increase *L* he suggested covering the copper wire continuously with either a thin envelope of soft iron or finely wound iron wire. This, he wrote, "should be of the greatest importance for long-distance telephony." Lorenz estimated that the method might increase the maximum speaking range by a factor of four.

The very same results were obtained by Heaviside, except that the Briton's analysis was more complete by taking into account the leakage factor from the beginning. Moreover, Heaviside investigated the possibility of two kinds of inductive "loading," both continuously and discretely by inserting induction coils in the line. Lorenz apparently did not consider the second possibility.

## 7. From engineering science to technology

The road from understanding scientifically how telephone currents propagate to the implementation of the knowledge in real telephone lines turned out to be long and bumpy. Heaviside's recommendation of adding induction coils was eventually transformed into an engineering theory with direct connection to practice, but it took more than a decade until the first loaded test cables were constructed and then proved their worth. The practical invention of coil loading was due to two American physicists working independently, Michael Pupin at Princeton University and George Campbell at the Bell System corporation. Their separate roads to the same technological result are well described in the historical literature [Britain 1970; Wasserman 1985]. After the Bell System had acquired Pupin's patent things went fast and long-distance telephony based on loading technologies became a reality latest by 1910.

Lorenz did not live to witness the success of inductively loaded cables. Interested in turning his theoretical discovery into a practical innovation, in late 1886 he addressed the large German cable manufacturer Felten & Guilleaume with a proposal of making a test cable in accordance with his theory. In a letter to the company he emphasized the economic benefits of changing from traditional cables to a new type where the copper conductor's self-induction was increased by covering it with a thin envelope of soft iron. "The usual fallacy that self-induction is a harmful agent to the propagation of telephone currents is a misconception," he pointed out; "if not corrected it will prevent the technicians from experimenting with copper wire covered with iron."

To obtain the same speaking distance without loading, the amount of copper would have to be greatly increased. According to Lorenz the loading method would save about 2,000 kg of copper for each 100 km. It is unknown how Felten & Guilleaume responded to Lorenz's arguments, which in part were scientific and in part economic, but in the end nothing came out of his contact with the German company. After 1886 Lorenz ceased working on telephone transmission and just shelved his calculations.

Lorenz's work was not entirely in vain, though, for it was known by a few Danish scientists and engineers. In 1901 Carl E. Krarup, a young telegraph engineer, made tests of a telephone cable of his own design but complying with Lorenz's method. During the next two decades such "Krarup cables" were manufactured commercially and widely used for submarine cable connections in particular [Kragh 1994]. The very first inductively loaded submarine cable was of Krarup's continuous type, laid in 1902 between Denmark and Sweden. The following year a 75 km Krarup cable connected Heligoland with mainland Germany and in 1911 the French Telegraph Company laid a cable of the same kind across the Channel. By that time more than fifty telephone cables of the design first suggested by Lorenz were in operation.

## 8. Conclusion

The work of the Danish physicist Ludvig Lorenz illustrates some general themes in nineteenth-century physics and in electromagnetic research in particular. The second part of the century was the period in which electromagnetic theory matured and also the period in which the theory was first applied to new technologies at a large scale. Lorenz was active in both areas. His electrodynamic theory of light based on equations involving retarded potentials has a notable position in the history of physics and the "Lorenz gauge condition" has become commonly known and adopted by modern physicists. As far as electrical technologies are concerned Lorenz's most remarkable work was possibly his theoretical analysis of the propagation of telephone currents a few years before Heaviside discovered the transmission equations. Since Lorenz did not publish this work, it played no direct role in the process that in the early years of the new century lead to inductively loaded cables. Lorenz was no Maxwell and no Lorentz either, but his work in electromagnetism is nonetheless of considerable historical interest.